\documentstyle[aps,epsf,epsfig,multicol]{revtex}

\begin{document}

\title{Scale-free Networks on Lattices}
\author{Alejandro F. Rozenfeld$^1$, Reuven Cohen$^1$, Daniel ben-Avraham$^2$, and Shlomo Havlin$^1$} 
\address{$^1$Minerva Center and Department of Physics, Bar-Ilan
University, Ramat-Gan 52900, Israel}
\address{$^2$Department of Physics, Clarkson University, Potsdam, New York 13699-5820}
\date{\today}
\maketitle

\begin{abstract}

We suggest a method for embedding scale-free networks, with degree distribution $P(k)\sim k^{-\lambda}$,
in regular Euclidean lattices.
The embedding is driven by a natural constraint of minimization of the total length 
of the links in the system. We find that all networks with $\lambda>2$ can be successfully
embedded up to an (Euclidean) distance $\xi$ which can be made as large as desired upon
the changing of an external parameter.  Clusters of successive 
chemical shells are found to be compact (the fractal dimension is $d_f=d$), 
while the dimension of the shortest path between
any two sites is smaller than one: $d_{min}=\frac{\lambda-2}{\lambda-1-1/d}$, contrary to
all other known examples of fractals and disordered lattices.

\end{abstract}

\begin{multicols}{2}

%%% INTRODUCTION

Many social, biological, and communication systems can be properly described  by complex
networks whose nodes represent individuals or organizations and links mimic the
interactions among them\cite{reviews}. An important class of complex networks are the
\emph{scale-free} networks, which exhibit a power-law connectivity distribution.
Examples of scale-free networks include the Internet\cite{internet,fal}, WWW\cite{bar_degree,broder_degree},
metabolic\cite{metabolic} and cellular networks\cite{cell}.
Most of the work done on scale free networks concerns off-lattice systems (graphs) where the
Euclidean distance between nodes is irrelevant.
However, real-life networks are often embedded in Euclidean space (e.g., the Internet is embedded
in the two-dimensional network of routers, neuronal networks are embedded in a three-dimensional 
brain, etc.). Indeed, in the case of the Internet, indications for the relevance of embedding space
is given in \cite{Bara_embedding}

In this Letter we develop a method for generating scale-free networks on Euclidean lattices
and study some of its properties. 
As a guiding principle we impose the natural restriction that the total length of links in the
system be minimal.

%%% THE MODEL IS DEFINED

Our model is defined as follows.
To each site of a $d$-dimensional lattice, of size $R$, and with periodic boundary conditions,
we assign a random connectivity
$k$ taken from the scale-free distribution 
\begin{equation}
P(k)=C k^{-\lambda} , \qquad m<k<K,
\label{Distr}
\end{equation}
where the normalization constant $C\approx(\lambda-1) m^{\lambda-1}$ (for $K$ large)\cite{Klafter}.
We then select a site at random and connect it to its closest neighbors until its (previously 
assigned) connectivity $k$ is realized, or until all sites up to a distance 
\begin{equation}
r(k)=Ak^{1/d}
\end{equation}
have been explored.
(Links to some of the neighboring sites might prove
impossible, in case that the connectivity quota of the target site is already filled.)
This process is repeated for all sites of the lattice.
We show that following this method networks with $\lambda>2$ can be successfully
embedded up to an (Euclidean) distance $\xi$ which can be made as large as desired upon
the changing of the external parameter $A$.

%%% DISCUSS k_c, xi

Suppose that one attempts to embed a scale-free network, by the above recipe, in an {\it infinite\/} lattice, $R\to\infty$.
Sites with a connectivity larger than a certain cutoff $k_c(A)$ cannot be realized, because of saturation of the 
surrounding sites.
Consider the number of links $n(r)$ entering a generic site from a surrounding neighborhood of radius $r$.
Sites at distance $r'$ are linked to the origin with probability $P(k'>(r'/A)^d)$: 
\begin{eqnarray}
P\left(k'>\left(\frac{r'}{A}\right)^d\right)=C\int\limits_{(\frac{r'}{A})^d} k^{-\lambda} dk
\\
\sim\left\{
\begin{array}{lc}
1 & r' < A. \\
(\frac{r'}{A})^{d(1-\lambda)} & r' > A.
\end{array}
\right.
\nonumber
\end{eqnarray}

Hence

\begin{eqnarray}
n(r)\sim\int \limits_{0}^{r} dr' r'^{d-1} P\left(k'>\left(\frac{r'}{A}\right)^d\right)
\\
\sim\frac{\lambda-1}{d(\lambda-2)}A^d-\frac{A^{d(\lambda-1)}}{d(\lambda-2)}r^{d(2-\lambda)}.
\nonumber
\end{eqnarray}

The cutoff connectivity is then

\begin{mathletters}
\begin{equation}
k_c=\lim_{r \to \infty} n(r)\sim\frac{1}{\lambda-2} A^d.
\label{kc}
\end{equation}
The cutoff connectivity implies a cutoff length
\begin{equation}
\xi=r(k_c)\sim (\lambda-2)^{-1/d} A^2.
\end{equation}
\end{mathletters}
The embedded network is {\it scale-free} up to distances $r<\xi$, and repeats itself (statistically) for
$r>\xi$, similar to the infinite percolation cluster above criticality: The infinit cluster in percolation
is {\it fractal} up to the coherence length $\xi$ and repeats thereafter \cite{havlin,book,stauffer}.

%%% DISCUSS r_max, K (R<infty)

When the lattice is finite, $R<\infty$, the number of sites is finite, $N\sim R^d$, which imposes
a maximum connectivity \cite{cohen,dor_cutoff}
\begin{mathletters}
\begin{equation}
K\sim m N^{1/(\lambda-1)}\sim R^{d/(\lambda-1)}.
\end{equation}
This implies a finite-size cutoff length
\begin{equation}
r_{max}=r(K)\sim A R^{1/(\lambda-1)}.
\end{equation}
\end{mathletters}
%%%%%DISCUSS   6 cases and show examples.
The interplay between the three length scales, $R$, $\xi$, $r_{max}$, determines the nature of the
network. If the lattice is finite, then the maximal connectivity is $k_{max}=K$ only if $r_{max}<\xi$.
Otherwise ($r_{max}>\xi$) the lattice repeats itself at length scales larger than $\xi$.
As long as $\min(r_{max},\xi)\ll R$, the finite size of the lattice imposes no serious restrictions.
Otherwise ( $\min(r_{max},\xi)\agt R$) finite-size effects become important.
We emphasize that in all cases the degree distribution (up to the cutoff) is scale-free.

\begin{figure}
%\centering
%\scalebox{1}{(a)}
%{\raggedright\raisebox{-2cm}{{\bf(a)}}}
\includegraphics[width=7cm]{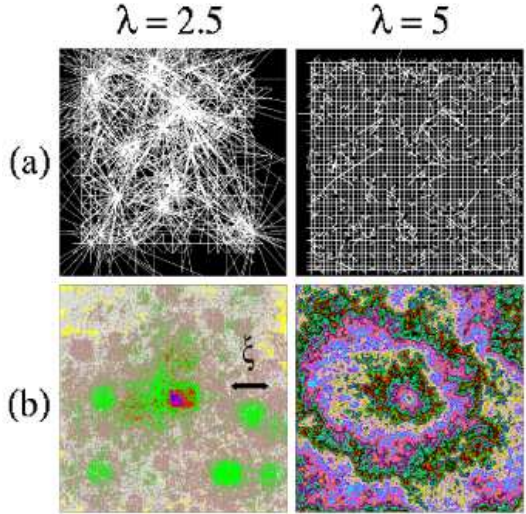}
%\raisebox{2cm}{{\bf(b)}}
%\raisebox{6cm}
%\hspace{7.5cm}\raisebox{2cm}{{\bf(b)}}
\caption{Spatial structure of connectivity network. (a) shown is the typical map
of links for a system of 50 x 50 sites generated from connectivity distributions with
$\lambda=2.5$ and $\lambda=5$. (b) shown are shells of equidistant
sites to the central one in a lattice of 300 x 300 sites. Note that for $\lambda=5$,
shells are concentric and continuous fractals; but for $\lambda=2.5$, shells are broken.} 
\label{pinta}
\end{figure}

In Fig. \ref{pinta}(a) we show typical networks that result from our embedding method, for $\lambda=2.5$
and $5$ in two-dimensional lattices (in this Letter, we limit our numerical results to $d=2$). 
The larger $\lambda$ is the more closely the network resembles the embedding lattice, because
longer links are rare~\cite{remark}. In Fig. \ref{pinta}(b) we show the same networks as in part (a) where successive
chemical shells are depicted in different shades.
Chemical shell $l$ consists of all sites at minimal distance (minimal number of connecting links) 
$l$ from a given site. 
For our choice of parameters, $\lambda=5$ happens to fall in the region of $\xi > r_{max}$, while for
$\lambda=2.5$, $\xi<r_{max}$. In the latter case we clearly see (Fig.\ref{pinta}(b), $\lambda=2.5$) 
the (statistical) repetition of the network
beyond the length scale $\xi$.
%\newline\raisebox{-2cm}{{\bf(a)}}
%\newline\raisebox{-6cm}{{\bf(b)}}
%\newline\hspace{0.1\textwidth}\raisebox{-0.5cm}{{\bf $\lambda=2.5$}}\hspace{0.15\textwidth}\raisebox{-0.5cm}{{\bf $\lambda=5$}}
%\newline\hspace{2cm}\raisebox{-0.5cm}{{\bf $\lambda=2.5$}}\hspace{0.15\textwidth}\raisebox{-0.5cm}{{\bf $\lambda=5$}}

\begin{figure}
\narrowtext
%\raggedright
%\raisebox{-2cm}{{\bf(a)}}
\raggedright
{\bf(a)}
\includegraphics[width=0.33\textwidth, angle=-90]{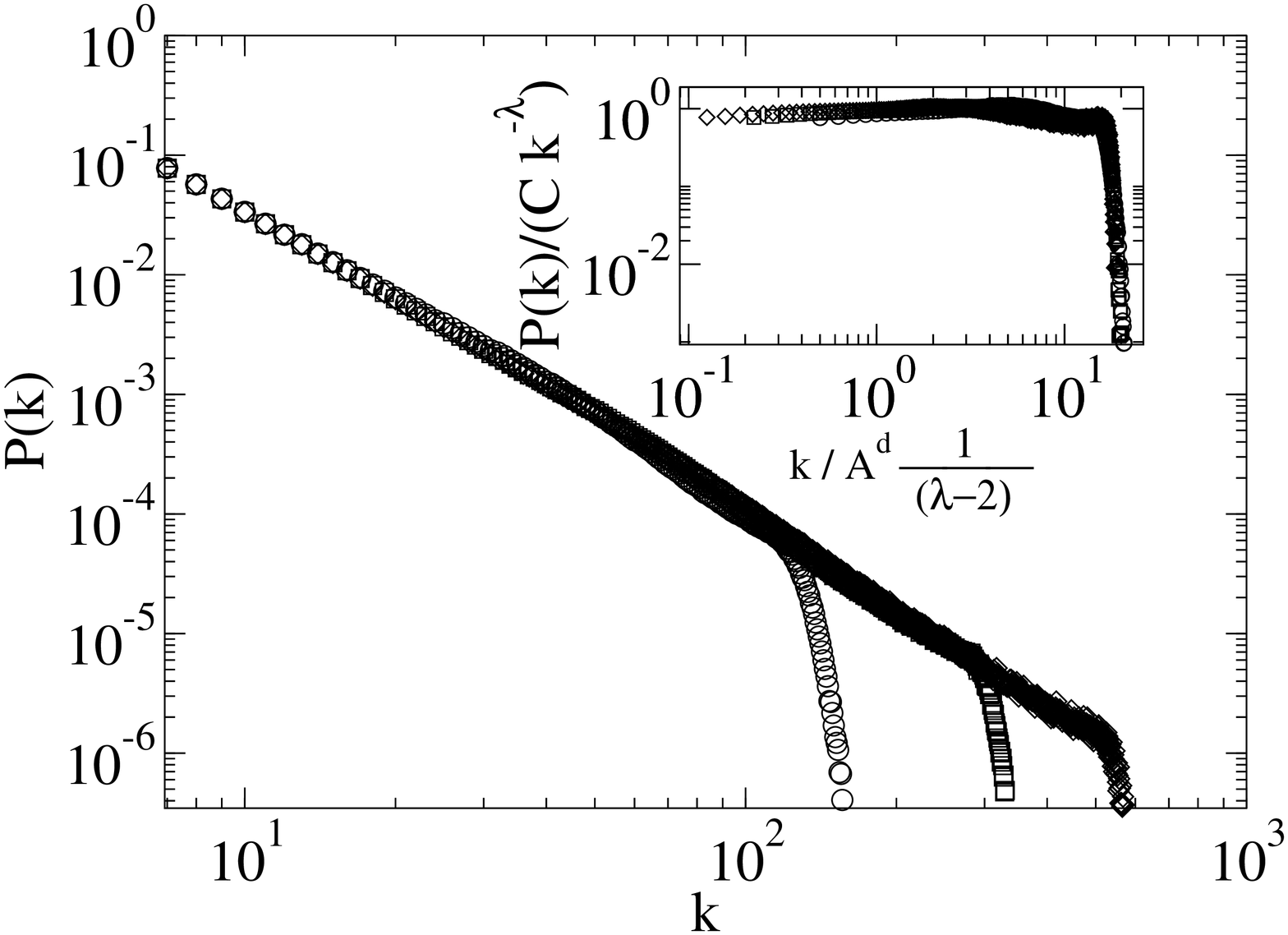} 
\\
%\raggedright
%\raisebox{-2cm}{{\bf(b)}}
%\raggedleft
{\bf(b)}
\includegraphics[width=0.34\textwidth, angle=-90]{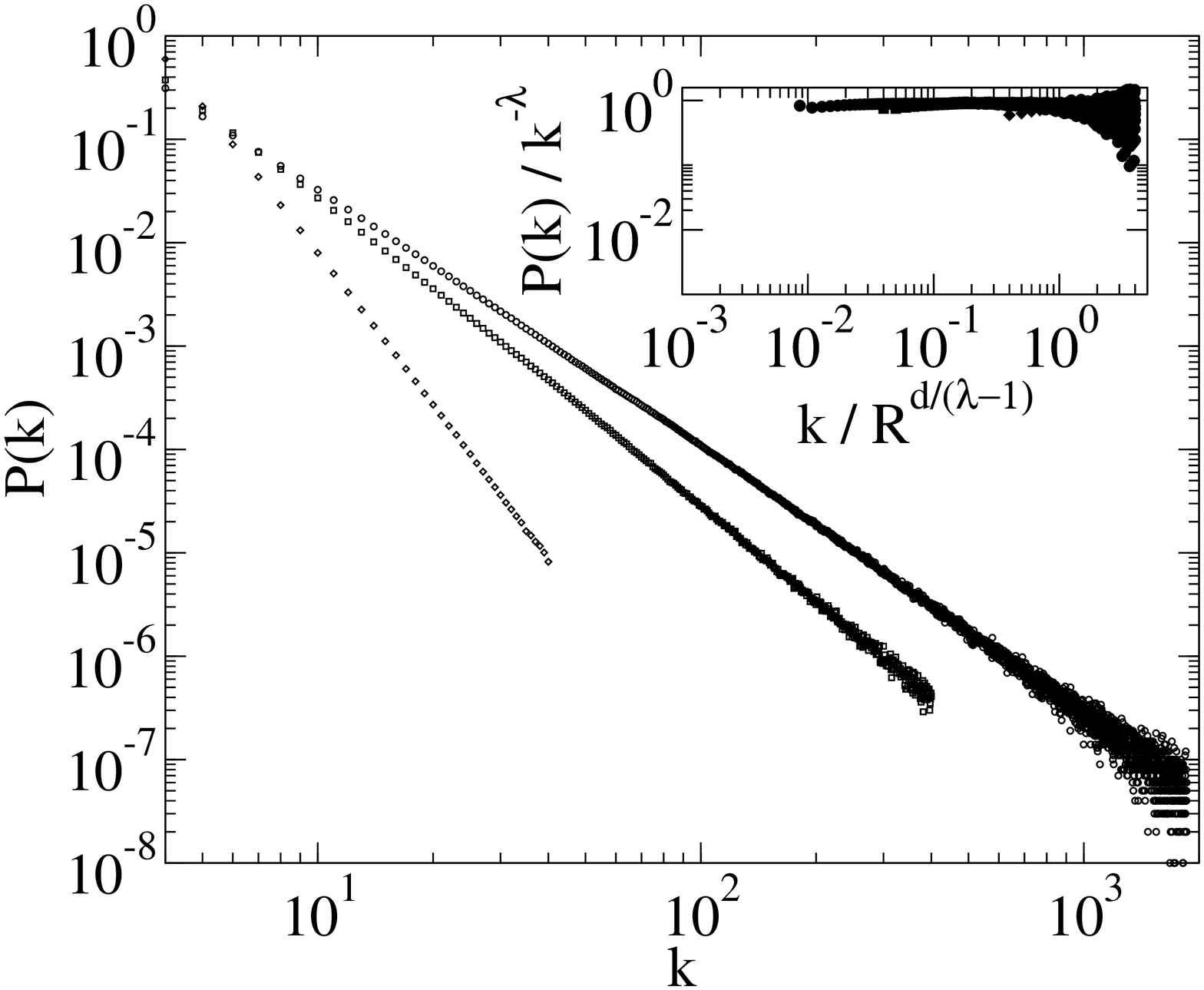} 

\caption{(a) The resulting connectivity distribution obtained from
simulations performed on two dimensional systems of size
$R=400$, $\lambda=2.5$ and for several values of $A$: (circles) A=2, (squares) A=3 and 
(diamonds) A=4; they all end at a cutoff $k_c(A)$. For this case $r_{max}>\xi$.
In the inset we show scaling collapse using same data. The threshold takes place at 
$k_c\sim A^d \frac{1}{\lambda-2}$ and confirms the validity of our theoretical estimations.
(b) Power law distribution of site connectivity in the network is showed for $R=100$, $A=10$ and
for different values of $\lambda$: $\lambda=2.5$ (circles), $3.0$ (squares), and $5.0$ (diamonds).
Note that in all cases the distribution achieves its (natural) cutoff $K$. In the inset we show
the corresponding collapse supporting $K\sim R^\frac{d}{\lambda-1}$. For this case, $r_{max}<\xi$.}
\label{fig:cutoff}
\end{figure}

The degree distribution resulting from our embedding method is illustrated in
Fig.~\ref{fig:cutoff}.  In Fig.~\ref{fig:cutoff}(a), $\xi<r_{max}$ and the distribution terminates at the cutoff $k_c$.
The scale-free distribution is altered slightly, for $k<k_c$, due to saturation
effects, but the overall trend is highly consistent with the original power-law.
The scaling in the inset confirms that $k_c\sim A^d$.  
In Fig.~\ref{fig:cutoff}(b), 
$\xi>r_{max}$ and the cutoff $K$ in the distribution results from the finite number of
sites in the system.  The scaling in the inset in Fig.~\ref{fig:cutoff}(b) confirms the known relation
$K\sim mR^{d/(\lambda-1)}$~\cite{cohen,dor_cutoff}.

\begin{figure}
\centering
\includegraphics[width=0.4\textwidth, angle=-90]{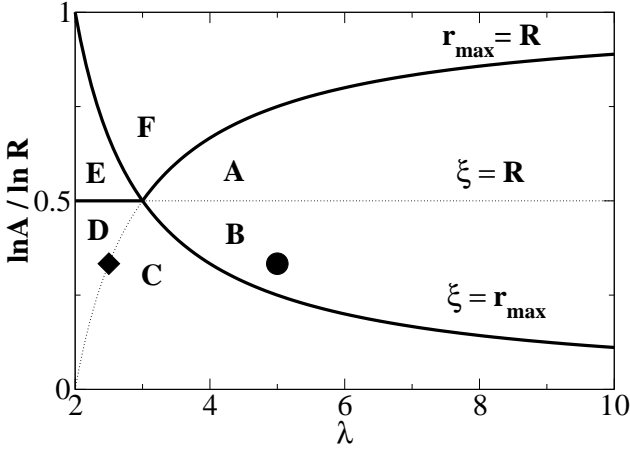} 
\caption{This diagram shows the six regions where different
behavior of the network is found: 
for region A: $r_{max}<R<\xi$, B: $r_{max}<\xi<R$, C: $\xi<r_{max}<R$, D: $\xi<R<r_{max}$,
E: $R<\xi<r_{max}$, F: $R<r_{max}<\xi$. The diagram can be mapped into only four
regions where the cutoff $k_c$ and where size effect $K$ are expected. A and B: no cutoff
and no size effect; C and D: cutoff and no size effect; E: cutoff and size effect; F:
no cutoff but size effect. The two symbols indicate the parameters corresponding to 
Fig. \ref{pinta}b, ( full diamond) $\lambda=2.5$ and (full circle) $\lambda=5$. }
\label{AvsAlfa}
\end{figure}

The different regimes are summarized in Fig. \ref{AvsAlfa}.

We now address the geometrical properties of the networks, arising from their embedding in
Euclidean space.  To this aim, it is useful to consider the spatial arrangement
of the networks as measured both in an Euclidean metric and in {\it chemical space}.
The chemical distance $l$ between any two sites is the length of the minimal path between them
({\it minimal\/} number of links).
Thus if the distance between the two sites is $r$, then $l\sim r^{d_{min}}$ defines the
minimal length exponent $d_{min}$.  We will see that $d_{min}<1$ (for $d>1$), contrary to all naturally
occurring fractals and disordered media.  Sites at chemical distance
$l$ from a given site constitute its $l$-th chemical shell. The number of (connected) sites within
radius $r$ scales as $m(r)\sim r^{d_f}$, defining the fractal dimension $d_f$.  Likewise, 
the number of (connected) sites within
chemical radius $l$ scales as $m(l)\sim l^{d_l}$, which defines the fractal dimension $d_l$ in
chemical space.  The two fractal dimension are related: $d_{min}=d_l/d_f$\cite{havlin,book,stauffer}.

\begin{figure}
\narrowtext
\includegraphics[width=0.4\textwidth,angle=-90]{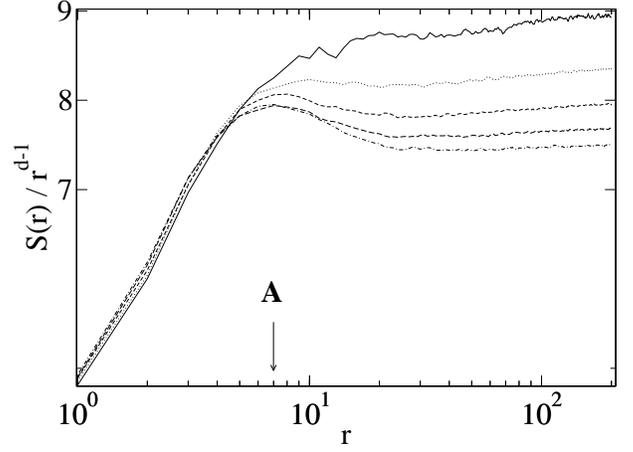} 
\caption{Plot of scaled perimeter as a function of the Euclidean distance from the central site, 
for several values of $\lambda$: $\lambda=3.0$ (top), $3.25$, $3.5$, $3.75$ and $4.0$ (bottom).
The simulations where performed with $A=7$. Note that the position where the curves split, $r\simeq A$,
is consistent with our analytical results (Eq. (\ref{Sr})). Also, the asymptotic
values shown for large $r$ are consistent with $c(\lambda)A$.}
\label{Perim}
\end{figure}

To study $d_f$, we compute the perimeter $S(r)$, the number of sites that connect the interior cluster
of a region of radius
$r$ to sites outside.  The fractal dimension then follows from the scaling relation $S(r)\sim r^{d_f-1}$.
We focus on the regime $\xi>r_{max}$.
Consider a shell $dr'$, of radius $r'$.  A site of connectivity $k'$ within the shell is connected to 
the outside (to a distance
larger than $r-r'$) with probability $P(k'>(\frac{r-r'}{A})^d)$, eq.~(3).  Thus,
\begin{eqnarray}
\label{Sr}
S(r)=\int_{0}^{r} dr'r'^{d-1} P(k'>(\frac{r-r'}{A})^d) \\
\sim\left\{
\begin{array}{lc}
r^d & r<A,\\ 
c(\lambda)Ar^{d-1} & r>A, 
\end{array}
\right.
\nonumber
\end{eqnarray}
where $c(\lambda)\sim 1+1/[d(\lambda-1)+1]$.  In other words, the network is compact, $d_f=d$ at large
distances $r>A$, and super-compact, $d_f=d+1$, at $r<A$.  Results for $d_f$ are presented in Fig.~\ref{Perim}
and are in good agreement with Eq.~(\ref{Sr}).  The slight slope observed for $r>A$ is due to 
analytical corrections, of order $r^{-1}$,
to the scaling $S(r)\sim r^{d-1}$, and can be obtained from a more careful analysis of
Eq.~(\ref{Sr}).

In order to compute $d_{min}$ (or $d_l$), we regard the chemical shells as being roughly smooth,
at least in the regime $\xi>r_{max}$, as suggested by Fig.~1b ($\lambda=5$). Let the 
width of shell $l$ be $\Delta r(l)$, then 
\begin{equation}
\label{l}
l=\int dl=\int \frac{dr}{\Delta r(l)}\sim r^{d_{min}},
\end{equation}
since $\Delta l=1$.  The number of sites in shell $l$, $N(l)$, is, on the one hand,
$N(l)\sim r(l)^{d-1}\Delta r(l)$.  On the other hand, since the maximal connectivity
in shell $l$ is $K(l)\sim N(l)^{1/(\lambda-1)}$, the thickness of shell ($l+1$) is
$\Delta r(l+1)$ which is determined by the length of the largest link to the next shell
i.e., $r[K(l)]$, and thus,
$\Delta r(l+1)\sim r[K(l)]\sim AK(l)^{1/d}$.  Assuming (for large $l$) that $\Delta r(l+1)\sim
\Delta r(l)$, we obtain
\begin{equation}
\Delta r(l)\sim r^{\frac{d-1}{d(\lambda-1)-1}}.
\end{equation}
Using this expression in~(\ref{l}), yields
\begin{equation}
\label{dmin}
d_{min}=\frac{\lambda-2}{\lambda-1-1/d}.
\end{equation}
Thus, above $d=1$, the dimensions $d_{min}$ and $d_l=d_{min} d_f$ are anomalous for all values of $\lambda$.

\begin{figure}
\narrowtext
\raggedright
%\raisebox{-2cm}{(a)}
{\bf(a)}
\includegraphics[width=0.4\textwidth,angle=-90]{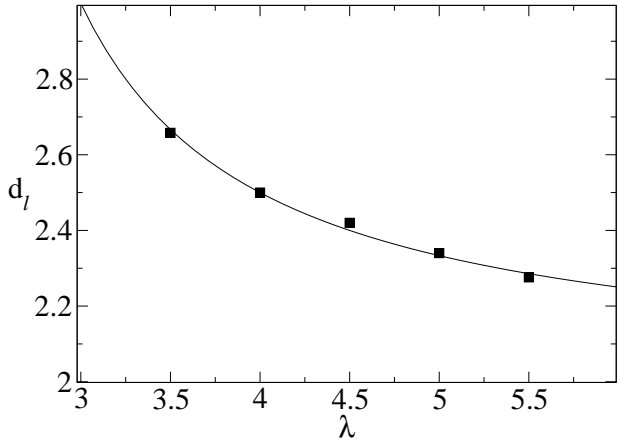} 
\\
%\raisebox{-2cm}{(b)}
{\bf(b)}
\includegraphics[width=0.4\textwidth,angle=-90]{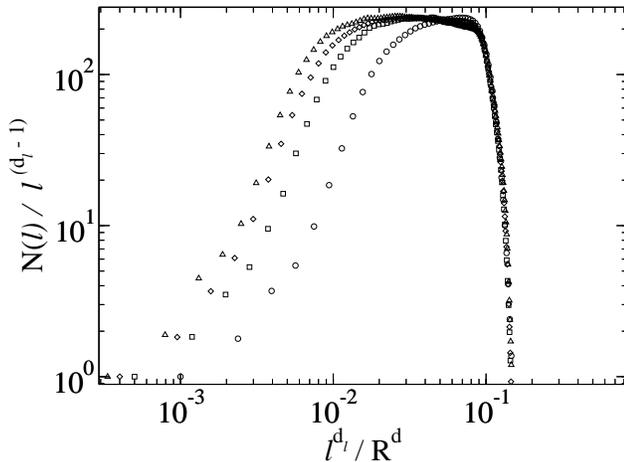} 

\caption{
{\bf(a)} The chemical dimension $d_l$ as a function of $\lambda$.
Note the good agreement between theoretical estimations (continuous line) and simulations
results (full squares).
{\bf(b)} The shape of the $\Phi(l^{d_l}/R^d)$ scaling function is shown for $\lambda=4$ and several   
lattice sizes: R=1000 (circle), 2000 (square), 2500 (diamond) and 3000 (triangle ). } 
\label{dlYscal}
\end{figure}

In Fig.~\ref{dlYscal}a we plot $d_l$ as measured from simulations, and compared with the analytical 
result Eq.~(\ref{dmin}). The scaling 
suggested in Fig.~\ref{dlYscal}b, $N(l)\sim l^{d_l-1}\Phi(l^{d_l}/R^d)$, is valid only for $\xi>r_{max}$.
For $R\to\infty$, we expect that the network is scale-free up to length scale $\xi$ and the analogous
scaling will be $N(l)\sim l^{d_l-1}\Psi(l^{d_l}/\xi^d)$,
where $\Psi(x\gg1)\sim x^{(d-d_l)/d_l}$.
\noindent

In summary, we propose a method for embedding scale-free networks in Euclidean lattices.
The method is based on a natural principle of minimizing the total length of links in the system.
This principle enables us to embed the scale-free in Euclidean space without additional external
exponents such as assumed by Manna and Sen\cite{Manna} and Xulvi-Brunet and Sokolov\cite{Sokolov}.
We have shown that while the fractal dimension $d_f$ of the network is the same as the Euclidean
dimension, the chemical dimension $d_l>d_f$ for all values of $\lambda$, yielding $d_{min}<1$ for all
$\lambda$ and $d>1$.

\end{multicols}

\begin{references}{}
\bibitem{remark} We choose $m=2d$ so that in the limit $\lambda\to\infty$ the network is
identical with the embedding lattice.  Clearly, this choice is not mandatory.
\end{references}
\end{document}